\documentclass[draft]{article}

 \newcommand{\beq}{\begin{equation}}
 \newcommand{\eeq}{\end{equation}}
 \newcommand{\beqa}{\begin{eqnarray}}
 \newcommand{\eeqa}{\end{eqnarray}}

 \def\<{\langle}
 \def\>{\rangle}

 \def\opone{\leavevmode\hbox{\small1\kern-3.8pt\normalsize1}}
 
 \newcommand{\complex}{{\kern .1em {\raise .47ex\hbox {$\scriptscriptstyle
 |$}}\kern -.4em {\rm C}}}
 \newcommand{\real}{{{\rm I} \kern -.19em {\rm R}}}

\def\K{{\mathbf k}}
\def\X{X}
\def\Y{Y}
\def\C{C}

\def\openone{\leavevmode\hbox{\small1\kern-3.8pt\normalsize1}}
\def\RR{{\rm I\kern-.2emR}}
\def\tr{{\rm tr}\; }

\def\ca{{\cal A}}

\def\GHZ{\ket{{\rm GHZ}}}

\newcommand{\ket}[1]{| #1 \rangle}
\newcommand{\bra}[1]{\langle #1 |}
\newcommand{\proj}[1]{\ket{#1}\!\bra{#1}}

\newcommand{\QED}{\hspace*{\fill}\mbox{\rule[0pt]{1.5ex}{1.5ex}}}

\newtheorem{definition}{Definition}
\newtheorem{observation}[definition]{Observation}
\newtheorem{theorem}[definition]{Theorem}
\newtheorem{proposition}[definition]{Proposition}

\newtheorem{corollary}[definition]{Corollary}

\begin{document}

\title{
Monotones and Invariants for Multi-particle Quantum States}
\author
{H Barnum${}^a$ and  N Linden${}^b$\\ 
{\protect\small\em ${}^a$Department of Computer Science,
University of Bristol,
}\\{\protect\small\em Merchant Venturers Building, Bristol BS8 1UB, UK}\\
{\protect\small\em ${}^b$School of
Mathematics, University of Bristol,
}\\{\protect\small\em University Walk, Bristol BS8 1TW, UK}
}
\date{28th March 2001}

\maketitle

\begin{abstract}
We introduce new entanglement monotones which generalize, to the
case of many parties, those which give rise to the
majorization-based partial ordering of bipartite states'
entanglement.  We give some examples of restrictions they impose
on deterministic and probabilistic conversion between multipartite
states via local actions and classical communication.  These
include restrictions which do not follow from any bipartite
considerations.  We derive supermultiplicativity relations between
each state's monotones and the monotones for collective processing
when the parties share several states. We also investigate 
polynomial invariants under local unitary
transformations, and show that a large class of these are
invariant under collective unitary processing and also multiplicative,
putting restrictions, for example, on the exact conversion of multiple
copies of one state to multiple copies of another.
\end{abstract}

\section{Introduction}
A key goal of quantum information theory is to understand the
local inter-convertibility of quantum states.  That is, given two
states, $\ket {\psi_1}$ and $\ket{\psi_2}$, we wish to find
conditions on $\ket {\psi_1}$ and $\ket{\psi_2}$ for one to be
converted into the other by local transformations.  Understanding
this issue is part of the more general question of characterising
what the truly different types of entangled quantum states are.

While much is known about the entanglement of bipartite quantum
states, multipartite entanglement appears to have a considerably
more complex structure. Many aspects of bipartite entanglement
have been fully understood in terms of a relation known as
majorization.  This relation gives necessary and sufficient
conditions for turning one pure state into another via local
operations and classical communication, and, when extended to
mixed states via a standard ``concave roof'' construction, gives
necessary and sufficient conditions for converting pure bipartite
states into mixed ones or ensembles of mixed ones, and necessary
conditions for general mixed-state conversion. Motivated in part
by the importance of multipartite entanglement to quantum
computation, in this paper we generalize these monotones to
multipartite systems, implying necessary conditions on multiparty
LOCC state transformations.

We also investigate aspects of polynomial invariants under local
unitary transformations, in particular their relevance to
collective processing by the relevant parties of several
multipartite states at once.

\section{Background: invariants and monotones}
A state is entangled if it cannot be prepared by initially
independent parties (each acting on one of the subsystems, or, as
we say, acting {\em locally}), even if these parties may
communicate classically.  (We use the standard acronym LOCC for
``local operations and classical communication''.)  More
generally, we may say one state is more entangled than another if
it cannot be prepared from the second state via LOCC.  Two states
are equivalent to one another, in terms of entanglement, if the
two states may be reversibly interconverted by LOCC.  One might
wonder if there exists a single measure of entanglement: a
function from states to the reals, such that a state may be
converted, by LOCC, into any state with an equal or lower value of
the function, but not to any state with a higher value of the
function.  The answer is no:  no single measure of entanglement
exists.  There are, however, many functions with the property that
no state may be converted to a state with a higher value of the
function---Vidal \cite{Vidal2000b} has dubbed these {\em
entanglement monotones}. Convertibility via LOCC is obviously a
partial order on the entangled states;  any proposed measure of
entanglement must be compatible with this partial ordering. In
particular, such a monotone must be an {\em invariant} under local
unitary transformations of the state.  The theory of {\em
polynomial}
invariants \footnote{That is, polynomial functions of the quantum state.} 
under actions of a group is particularly
interesting, and although such invariants are not (at least not
{\em prima facie}) guaranteed to be entanglement monotones, the
close connection between local unitary invariance and
entanglement, and the mathematical importance of polynomial
invariants, suggest that much may be learned about entanglement by
studying the equivalence classes of states having fixed values of
the polynomial invariants \cite{Linden98a}.  For example, any
bipartite entanglement monotone must be, on  pure states
$\ket{\psi^{12}}$, a
function solely of the eigenvalues $\lambda_i$  of the reduced  density
operator $\rho := \mbox{tr}_2 \proj{\psi^{12}},$ 
these being invariant under local unitaries.  And these
eigenvalues may be recovered as the solutions of the system of $d$
polynomial equations in the variables $\lambda_i$, $i=1,..,d$
\beqa \sum_{i=1}^d \lambda_i^k = X_k, ~~~~~~k=1,...,d\;, \eeqa
where $X_k := \mbox{tr } \rho^k$, 
are $d$ polynomial invariants, each homogeneous
of degree $k$. 

Some of the polynomial invariants are themselves
entanglement monotones. 
In fact, the $d$ polynomial invariants
just defined are increasing entanglement monotones:
they increase or stay
constant under LOCC.  This may be proved using Example II.3.5(iii) in
\cite{Bhatia97a}.
These are
not, however, complete (by complete we mean that their nondecrease is a
necessary and sufficient condition for pure-state to pure-state
transitions with certainty). 
To see this,  consider for example
states $\ket{\psi_1}$ with reduced density matrix
eigenvalues
$.5, .3, .2,$ and $\ket{\psi_2}$ with reduced density matrix eigenvalues
$.51, .28, .21$.  These have $(X_1, X_2, X_3) = (1, .38, .16)$ and 
$(1, .3826, .163864)$ respectively, so these invariants are nondecreasing
as $\ket{\psi_1} \rightarrow \ket{\psi_2}$.  This is necessary for 
$\ket{\psi_1} \rightarrow \ket{\psi_2}$ with certainty via LOCC, and if it were 
sufficient, that transition would be possible.  However, 
the vector $(.51, .28, .21)$ does not majorize $(.5, .3, .2)$, and this 
majorization is known to be necessary (cf. below) for the transition
in question.

Also, the elementary symmetric polynomials $S_k$ in the eigenvalues of the 
reduced density matrix, as well as the ratios $S_k/S_{k-1}$
of them, are (increasing) entanglement monotones (the proof uses 
Example II.3.16 and Exercise 
II.3.19 in \cite{Bhatia97a}).

In quantum mechanics any measure of how mixed a density operator
is can be converted into a candidate 
measure of how entangled a pure bipartite
state is.  The mixedness of the reduced density operator might be thought to
measure the
entanglement of the state.  For example, the reduced density matrix
entropy is one common measure \cite{Bennett96a};  another is the trace of the
square of the reduced density matrix.  Alberti and Uhlmann 
\cite{Alberti81a}, as well as Wehrl \cite{Wehrl74a, Wehrl78a} and others,
have extensively studied a partial ordering $\succeq$ of density matrices:
$\rho \succeq \sigma,$ read ``$\rho$ is more mixed than $\sigma,$'' 
if $\rho$
is a convex combination of unitary transforms of $\sigma$:
\beqa \label{mixed state ordering}
\rho = \sum_i p_i U_i \sigma U_i^\dagger\;.
\eeqa
This can be shown to be equivalent to the statement that the
vector $\lambda(\rho)$ 
whose components are $\rho$'s eigenvalues arranged in decreasing
order is majorized by the vector of $\sigma$'s decreasingly
ordered eigenvalues.  An important fact about majorization is that
if a vector $\lambda$ majorizes a vector $\mu$, $\mu$ may be 
obtained by multiplying $\lambda$ by a doubly stochastic matrix (one whose
rows and columns sum to unity).  Birkhoff and von Neumann showed that
any doubly stochastic matrix is a convex combination of permutation matrices.

It is therefore natural to require that any reasonable measure of
entanglement be compatible with the partial ordering ``$\ge$'' 
on pure states, defined by:
\beqa \label{pure state ordering}
\ket{\psi^{12}} \ge  \ket{\phi^{12}} :=
\rho_\psi^1 \succeq \rho_\phi^1\;.
\eeqa
(Here $\rho_\psi^1$ and $\rho_\phi^1$ are the reduced density matrices of
the states.)
Any reasonable measure of entanglement should also satisfy that it
not increase under LOCC.
If all such measures must be compatible with the above partial
ordering, then it must be impossible, by LOCC, to go from one
pure bipartite state to another more entangled than it according
to the ordering (\ref{pure state ordering}).  Indeed, since the ordering
(\ref{mixed state ordering}) appears to be the whole story about whether one
density operator is more mixed than another, it was also natural to conjecture
that the ordering (\ref{pure state ordering}) 
is the full story with respect to whether one pure bipartite
state is more entangled than another.  In operational terms, this means
that the condition 
$\ket{\psi^{12}} \ge \ket{\phi^{12}}$ is not only necessary, but also sufficient
for converting the $\ket{\psi^{12}}$ to $\ket{\phi^{12}}$ 
by LOCC, and that there should be a 
protocol to do this conversion by making 
use of the Birkhoff-von Neumann decomposition in some simple way. 
This conjecture was proved by Nielsen \cite{Nielsen99a}.   While Nielsen's protocol
and Hardy's \cite{Hardy99a} version of it are clearly closely related to the Birkhoff-von
Neumann decomposition, Jensen and Schack's version 
\cite{Jensen2000b} uses it most directly.

Nielsen's result implies that for pure states on the tensor product 
$\mathbf{C}^d \otimes \mathbf{C}^d$,
the $d$ quantities:
\beqa
E_k(\ket{\psi^{12}}) := \sum_{i=1}^k \lambda_i^\downarrow(\rho^1)
\eeqa
are {\em  entanglement monotones}.  For these quantitities cannot decrease under LOCC.

The term {\em entanglement monotones} is sometimes reserved for quantities
which cannot increase under LOCC;  here we allow either nonincreasing, or nondecreasing,
monotones;  they are equally useful, as 
it is trivial to obtain one of one type from one of the other.  For convenience,
we will call a monotone which cannot decrease under LOCC an {\em increasing}
entanglement monotone, and one which cannot increase under LOCC a {\em decreasing}
entanglement monotone.

In the remainder of this paper, we will investigate some generalizations
of the majorization-derived monotones, and of 
some polynomial invariants under local unitaries, 
to multipartite systems.  In particular, we will examine multiplicativity of such quantities
when a given set of parties has several, independent, shared states upon which they
may operate.  While the generalizations of the majorization-monotones will 
be supermultiplicative
(their multiplicativity remaining an open question), 
some cases for which multiplicativity holds
will be investigated.  A large class of the polynomial invariants will, on the other hand, be 
shown to be multiplicative.  Multiplicativity is an important property in investigating
transformations between many copies a given state, both for a finite number of 
copies and in the asymptotic limit in which the 
rate of conversion of multiple copies
of one state into another
is of interest.

\section{Multipartite monotones}
\subsection{Definition of the monotones}
How might we generalize the majorization-derived monotones 
to multipartite systems?  There is a
 well-known variational
characterization \cite{Horn85a} of the sums $\Sigma_k$ 
of the $k$ largest eigenvalues of an operator
as $\Sigma_k(\rho) = 
\max_{{\rm\scriptscriptstyle rank-{\it}k~projectors~ P}} \tr P \rho$.
We may use this to characterize the bipartite
quantities $E_k$ by
\beqa
E_k(\ket{\psi^{12}})\equiv 
\max_{{\rm\scriptstyle rank-{\it k}~projectors {\it P}}} 
|| I \otimes P \ket{\psi^{12}}||^2.
\eeqa
We propose to generalize this definition of the bipartite quantities to multipartite 
systems, in the following way.
\begin{definition}
For an $N$-partite quantum system in a (not necessarily normalized)
pure state $\ket{\psi^{123...N}}$, 
define 
\beqa \label{themonotones}
E_{k_1,k_2,...,k_N}(\ket{\psi^{123...N}}) := 
\max_{\Gamma_1,...,\Gamma_N} ||\Gamma_1 \otimes \cdots \Gamma_N
\ket{\psi^{123...N}}||^2\;,
\eeqa
where each of 
$\Gamma_i$ is a $k_i$-dimensional projector in system $i$.
\end{definition}
This is the squared norm of the maximal projection of the state
onto a tensor product of local subspaces 
having dimensions $k_1,...,k_N$.  The integers $k_i$ may range
from $1$ to $d_i$, the dimension of the $i$-th party's Hilbert
space.  To reduce clutter we will sometimes
write the ``multi-index'' $\K$ for $k_1, ..., k_N.$  For every 
$E_\K$, its maximal value on the set of pure states with squared norm
equal to $X$ is just $X$;  thus the maximal value
for normalized states is $1$, and is attained on pure product states.
Note
that some non-product states may also have $E_\K=1$ for some $\K$.
However, only product states can have $E_{111...1} = 1$.

\begin{observation}
$E_{k_1,...,k_N}(\ket{\psi^{1,...,N}})$ are 
invariant under local unitary transformations
of $\ket{\psi^{1,...,N}})$.
\end{observation}
This is immediate from (\ref{themonotones}).  
Explicitly
\beqa
E_{k_1,...,k_N}(U_1 \otimes \cdots \otimes U_N \ket{\psi^{123...N}}) = 
\nonumber \\
\max_{\Gamma_1,...,\Gamma_N}\bra{\psi^{123...N}} U_1^\dagger \otimes \cdots \otimes U_N^\dagger  
(\Gamma_1 \otimes \cdots \Gamma_N)
U_1 \otimes \cdots \otimes U_N \ket{\psi^{123...N}}
\nonumber \\
= 
\max_{\Gamma_1,...,\Gamma_N}
\bra{\psi^{123...N}}
(U_1^\dagger \Gamma_1 U_1\otimes \cdots \otimes U_N^\dagger 
\Gamma_N U_N)
\ket{\psi^{123...N}}\;.
\eeqa
For any set of $\Gamma_i$ and initial state $\ket{\psi}$
the same value of the maximand will be achieved with the 
local-unitarily transformed state and the projectors transformed
by the inverse;  these are also projectors of the same rank, 
so the maximum over all local projectors 
is the same for both initial states. \QED

One might consider the analogous definition, but with $\Gamma_i$
replaced by
rank-$k_i$ partial isometries.  A rank-$k$ partial isometry
may be written as $\Gamma U$, where $U$ is unitary and $\Gamma$ is
a rank-$k$ projector.  Therefore this 
would define the same quantities, by the same argument just
used for unitary invariance.   
Explicitly, this definition would run:
\beqa \label{isometriesform}
\tilde{E}_{k_1,k_2,...,k_N} := 
\max_{R_1,...,R_N} ||R_1 \otimes \cdots R_N
\ket{\psi^{123...N}}||^2\;,
\eeqa
where $R_i$ are partial isometries with ranks $k_i$.
 

We extend these monotones to mixed states via the usual 
``concave roof'' device of defining
the mixed state quantity to be the maximum
of the average of the pure state quantity, over ensembles of pure
states for the mixed state in question:
\beq
E_{k_1,...,k_N}(\rho) := \max_{\{\ket{\psi_i}\}_i: \sum_i \proj{\psi_i} = \rho}
\sum_i E_{k_1,...,k_N}(\ket{\psi_i})\;.
\eeq
For normalized mixed states, the maximum 
of $E_\K$ is achieved at separable states, with $E_\K=1$.
Again, while some entangled states may have some $E_\K=1$, 
only separable states can have $E_{111..11}=1.$

Extend each monotone $E_\K$
to {\em ensembles} $\{p_i,\rho_i\}_i$ of states via
$E_\K(\{p_i,\rho_i\}_i) := \sum_i p_i E_\K(\rho_i)$.
Note 
that the monotones are linearly homogeneous in the density
operators:
\beq
E_\K(\lambda \rho) = \lambda E_\K(\rho)\;.
\eeq
This has the consequence that we may 
represent an ensemble $\{p_i, \rho_i\}_i$ by the subnormalized
operators $\tilde{\rho}_i$, with $\tr \tilde{\rho}_i = p_i$,
and then the ensemble average of the monotone
is just $\sum_i E_\K(\tilde{\rho_i})$.
This representation of ensembles by sequences of unnormalized
density operators is useful because the arguments we use will often involve 
successively finegraining an ensemble, in which case it
is slightly cumbersome to renormalize and keep track of the 
probabilities introduced at each step.

Throughout the paper we use a notation in which 
sets or ensembles may be referred to by expressions with
curly braces around them, such as $\{\ket{\psi_{ij}}\}_i$.  
Some indices in the expression within
braces ($i$, in our example) 
also appear as subscripts of the right-hand brace: this indicates
that the set consists of all the values taken by
the expression within the
braces as these indices vary.
If there are also ``free'' indices (like $j$ in our example)
in the 
expression within the braces, which don't appear as subscripts 
of the right-hand brace, 
the overall expression including braces and subscripts 
ranges over 
{\em different} sets or ensembles as these free indices vary.
Thus $\{\ket{\psi_{ij}}\}_i$ refers to the $j$-th ensemble of 
some set of ensembles indexed by $j.$  Each of these ensembles
consists of the states $\ket{\psi_{ij}}$, for all values of $i$.
When we view this as an ensemble, we take the probability for 
each state to be given by its squared norm $||~\ket{\psi_{ij}}||^2.$  
The point of this notation is just to make it clear, when we
are considering many ensembles at once, which indices identify
the ensemble, and which identify the states within each 
ensemble.  (A similar notation is sometimes used by mathematicians
to specify matrices by their matrix elements or tensors by their
components, to distinguish indices specifying which component
of a tensor from indices specifying
which tensor.)

The following observation partly explains the terminology
``concave roof.''
\begin{observation}[Concave roofs are concave] 
\label{obs: concave roofs are concave}
Let 
\beq\label{a very important definition}
R(\rho) := \max_{\{\ket{ \psi_i}\}_i : \sum_i \proj{\psi_i} = \rho}
\sum_i Q(\ket{\psi_i})\;.
\eeq
Then $R(\rho)$ is concave in $\rho$, i.e. 
\beq
\sum_k R(\rho_k) \le R(\sum_k \rho_k)\;.
\eeq
\end{observation}
{\em Proof:~~} Define $Q$ on pure-state
ensembles by $Q(\{\ket{\psi_i}\}_i) := \sum_i Q(\ket{\psi_i})$.
We do the case $k\in\{1,2\}$; the general case
follows by a trivial induction or by the same proof with 
wider-ranging indices.
Consider states $\rho_1$, $\rho_2,$ and pure-state ensembles
$\Upsilon_1=\{\ket{\psi^1_i}\}_i$ and 
$\Upsilon_2=\{\ket{\psi^2_i}\}_i$ for $\rho_1$, 
$\rho_2$ respectively.  The ensemble made from the states of
both ensembles,
$\Upsilon := \{\ket{\psi^k_i}\}_{k,i},$ is a pure-state ensemble for 
$\rho_1 + \rho_2$.  Now, $Q(\Upsilon) = Q(\Upsilon_1) 
+ Q(\Upsilon_2)$ from the definition, so if $\Upsilon_1$ and 
$\Upsilon_2$
achieve the maximum in (\ref{a very important definition}),
then $Q(\Upsilon) := R(\rho_1) + R(\rho_2).$  But as $\Upsilon$ is
a pure-state ensemble for $\rho := \rho_1 + \rho_2$, by 
(\ref{a very important definition}) $R(\rho)$ cannot
be less than $Q(\Upsilon)$.  \QED

To help the reader get 
used to our notation for ensembles, using unnormalized
states, we
record for comparison the more standard way of writing 
concavity, with ensembles of normalized states
and explicit probabilities:
\beq
\sum_i p_i R(\hat{\rho}_i) \le R(\sum_i p_i \hat{\rho}_i)\;.
\eeq

It follows from Observation \ref{obs: concave roofs are concave} 
that $E_{k_1,...,k_N}$ are concave.
$E_{k_1,...,k_N}$ are candidates for (increasing) multipartite entanglement monotones.
They generalize the bipartite case.

\begin{proposition} \label{prop: bipartite}
For a bipartite system, pure states $\ket{\psi^{12}}$
satisfy:
\beqa
E_{k_1,k_2}(\ket{\psi^{12}}) = E_{k_2,k_1}(\ket{\psi^{12}}) 
= \sum_{i=1}^{\min(k_1,k_2)} \lambda_i(\rho_1)\;.
\eeqa
\end{proposition}
(Recall that $\lambda_i(\rho)$ is the $i$-th decreasingly ordered
eigenvalue of $\rho$.)
The proof makes interesting use of some tools which are useful
in many places in quantum information theory;  to avoid interrupting
the flow of our exposition, it appears 
in an Appendix.

\subsection{Demonstrating monotonicity}

Vidal \cite{Vidal2000a} gave succinct
necessary and sufficient conditions for a
quantity to be an increasing entanglement monotone.  
Such a quantity must be {\em concave}, and 
{\em increasing under unilocal operations}.
Concavity means that if we throw away information 
about which state of the ensemble we have, the expected
entanglement decreases (the monotone increases):
\beqa
M(\{\ket{\psi_{ij}}\}_{ij}) \le \sum_i M(\{\sum_j \proj{\psi_{ij}}\}_i)\;.
\eeqa
Increase under unilocal operations
means that under any set of quantum
operations $\ca_m$ on one subsystem
which sum to a trace-nonincreasing operation
$\ca = \sum_m \ca_m$, we have:
\beqa
M(\rho) \le \sum_m M(\ca_m(\rho))\;.
\eeqa

(Each of the $\ca_m$ is assumed to act nontrivially on the 
{\em same} party's subsystem, and only on that subsystem.)
We have already shown, in Observation \ref{obs: concave
roofs are concave},
that $E_{k_1,...,k_N}$ are concave.
They are also increasing under unilocal operations:
\begin{proposition}\label{monotonicincreasing}
$E_{k_1,...,k_N}$ increases or remains constant under
unilocal quantum operations.
\end{proposition}

{\em Proof:}
The most general objects on which our monotones $E$
are defined are ensembles of (possibly mixed) states. 
Consider an ensemble of unnormalized states $\rho_i$;
then $E_\K(\{\rho_i\}_i) := \sum_i E_\K(\rho_i)$.
To show that this increases under unilocal operations,
it suffices to show that $E_\K(\rho_i)$ does for each $i$.
We therefore suppress the index $i$ and consider 
a single mixed input state $\rho$.
Let $\{\ca_m\}_m$ be a unilocal set of operations, which we take
WLOG to be on system $N$.
We wish to show that 
\beq
E_\K(\rho) \le \sum_m E_\K(\ca_m(\rho))\;.
\eeq
Each $\ca_m$ has a Hellwig-Kraus decomposition $A_{ms}$ with 
$s$ taking a finite number of values, so
that 
$\ca_m(\rho) = \sum_{s} A_{ms} \rho A_{ms}^\dagger$.
By concavity,
\beqa\label{an equation}
\sum_{ms} E_\K(A_{ms} \rho A_{ms}^\dagger) \le 
\sum_m E_\K(\sum_s A_{ms} \rho A_{ms}^\dagger)\;.
\eeqa
Therefore, we will show monotonicity under sets $\{\ca_j\}_j$
of ``one-operator'' unilocal operations,
for which each $\ca_j$ corresponds to a single operator $A_j.$
For, taking $j$ to be the double index $ms$ in (\ref{an equation}),
monotonicity under sets of general unilocal operations will follow 
from monotonicity under sets of one-operator unilocal operations
and (\ref{an equation}).
To show monotonicity under one-operator sets of unilocal operations, 
let 
$\{\ket{\psi^{max}_i}\}_i$ be a pure-state ensemble for $\rho$ 
which achieves the maximum in the convex roof expression 
\beq
E_{k_1,...,k_N}(\rho) := \max_{\{\ket{\psi_i}\}_i: \sum_i \proj{\psi_i} = \rho}
\sum_i E_{k_1,...,k_N}(\ket{\psi_i})\;.
\eeq
Under unilocal operations $\ca_j$ each with one Hellwig-Kraus operator
$A_j$ on the $N$-th system, we have 
\beq
\rho \rightarrow \{A_j \rho A_j^\dagger\}_j =: 
\{\rho_j\}_j\;.
\eeq
Suppose that 
$\hat{\Gamma}^i_1,...,\hat{\Gamma}^i_{N}$ achieve the maximum,
for the initial state $\ket{\psi^{max}_i},$
in the definition (\ref{themonotones}) of $E_\K.$
Define
\beq \label{defn psitilde}
\ket{\tilde{\psi}_i} := \hat{\Gamma}^i_1 \otimes \hat{\Gamma}^i_2
\otimes \cdots \hat{\Gamma}^i_{N-1} \ket{\psi^{max}_i}\;.
\eeq
Then 
\beq \label{la dee dah}
E_{k_1,...,k_N}(\ket{\tilde{\psi}_i}) = 
\max_{\Gamma_N^i} \left|\Gamma_N^i \ket{\tilde{\psi}_i}\right|^2
 = E_{k_N}(\ket{\tilde{\psi}'_i})
\eeq 
where $\ket{\tilde{\psi}'_i}$ is $\ket{\tilde{\psi}_i}$
considered as the state of a bipartite 
system in which  systems $1,...,N-1$ are viewed as
a single quantum
system, tensored with system $N$.  
By the definition (\ref{defn psitilde}) of $\ket{\tilde{\psi}_i}$,
\beqa
E_\K(\rho) = \sum_i E_{k_1,...,k_N}(\ket{\tilde{\psi}_i}) \;.
\eeqa
By (\ref{la dee dah}), this is equal to
\beq 
\sum_i E_{k_N}(\ket{\tilde{\psi}'_i})
\le \sum_{ij} E_{k_N}(A_j \ket{\tilde{\psi}'_i}) \;,
\eeq
where the inequality is due to the monotonicity of the 
bipartite monotones $E_k$ \cite{Vidal99a}.
Inserting the definition of these bipartite monotones, 
the right-hand side is 
\beq
\sum_{ij} \max_{\Gamma_N^{ij}} \left| \Gamma_N^{ij} 
A_{j} \ket{\tilde{\psi'_i}} \right|^2\;.
\eeq
Using the definitions of $\ket{\tilde{\psi}'_i}$ 
and $\ket{\tilde{\psi}_i}$ gives:
\beq
\sum_{ij} \max_{\Gamma_N^{ij}} \left| 
\hat{\Gamma}_1^{i} \otimes \cdots \otimes \hat{\Gamma}_{N-1}^{i} 
\otimes\Gamma_N^{ij}
A_j \ket{\psi^{max}_i} \right|^2 
\eeq
which
is less than or equal to  
\beqa
\sum_{ij} \max_{\Gamma_1^{ij},...,\Gamma_N^{ij}} \left| 
{\Gamma}_1^{ij} \otimes \cdots \otimes {\Gamma}_{N-1}^{ij} 
\otimes\Gamma_N^{ij}
A_j \ket{\psi^{max}_i} \right|^2 \nonumber \;,
\eeqa
(with $\Gamma^{ij}$ constrained to 
have the ranks $k_i$)
since we have just widened the domain of maximization.
By the definition of $E_{\K}$, this is equal to 
\beqa
\sum_{ij} E_\K (A_j \ket{\psi_i^{max}}) =
E_\K(\{A_j\ket{\psi^{max}_i}\}_{ij})\;.
\eeqa
We now note that $\{A_j \ket{\psi_i^{max}}\}_i$ (for fixed $j$)
is an ensemble for $\rho_j$, hence
by the concavity of $E_\K$
\beq
E_\K(\{A_j\ket{\psi^{max}_i}\}_{ij})
\le E_\K(\{\rho_j\}_j)\;,
\eeq
 as required.
\QED

The conjunction of Proposition 5 and the concavity of $E_{\K}$ gives
\begin{theorem}
$E_{\K}$ are entanglement monotones.
\end{theorem}

When analyzing particular multipartite states, we should remember
that this definition of monotones gives us not only the monotones 
explicitly mentioned in Definition \ref{themonotones}, 
but also all the monotones given by the same definition, but with
some subsets of the set of systems grouped and considered as 
single systems, and Definition \ref{themonotones} applied to 
this ``coarsegrained'' party structure.   These are also 
monotones under LOCC with respect to the finegrained party structure,
since operations local with respect to the finegrained structure
are also local with respect to the coarser one.
(Some obvious inequalities therefore 
hold between monotones and coarse-grainings
of them.)
An example of this construction is the frequent practice of 
grouping the parties into two disjoint sets, and applying bipartite 
monotones to the resulting bipartite structure, when studying 
multipartite states.  Note, however, that while our multipartite 
monotones include all such bipartite monotones based on majorization
of the reduced density matrix of some set of parties, they also 
include, as we will show in Section \ref{sec: examples}, 
irreducibly multipartite monotones
giving us information not provided by the majorization-based monotones
studied by Nielsen and Vidal.

\section{Collective processing and multiplicativity} \label{collective monotones}
Suppose we have two multiparty states, $\ket{\psi^\X}$ and $\ket{\chi^\Y}$,
on Hilbert spaces $\X$ and $\Y$ each composed of subsystems $1,...,N$
held by $N$ different parties.  
Thus $\X = \X^1 \otimes \X^2 \otimes \cdots \otimes \X^N$,
and similarly for $\Y$,  but there is no assumption that $\X^i$ and $\Y^i$ are 
isomorphic.  For the purposes of multipartite LOCC protocols, party $i$ may operate
on 
his part of both the $\X$ and the $\Y$ systems; in other words,
{\em arbitrary} operations by $i$ 
on 
$\X^i \odot \Y^i$ are considered local operations in this framework.
\footnote{If we have two states $\rho_1$ and $\rho_2$ we
 will denote their tensor product $\rho_1\odot \rho_2$;
 the aim of this notation is to reinforce the fact that this is
 not the tensor product of the Hilbert spaces of the  parties
 (for which we will use the usual notation $\otimes$).}
We will refer to such operations, and their obvious generalization
to more than two states shared by the same set of parties,
as {\em collective} processing of
the states (of $\ket{\psi^\X}$ and $\ket{\chi^\Y}$, in the two-state case). 

 The question of collective processing is important for a number
 of reasons.  Firstly it is known that collective processing is
 needed, in general, to distill entanglement from quantum states.
 Specifically, given a quantum state, one may wish to perform (LOCC)
 operations on it so that there is some probability that one of
 the outcomes of the measurements has more entanglement than the
 initial state (of course one cannot increase the average
 entanglement of the outcomes).  It has been shown however
 \cite{Linden98b} that, in general, entanglement
 distillation is not possible if one only has one copy of a qubit
 state, even though with collective processing of more than one copy of
 the same state, distillation is possible.  Secondly in order to
 extract the maximum possible entanglement from a pure bipartite
 state (i.e. the local entropy of one of the parties) one needs
 collective processing \cite{Bennett96c}.  

When using monotones or invariants to investigate
collective processing, it is important to know
about the behavior of the quantities in question under the tensor 
product $\odot$
of different states of the same parties.  If the quantities are
additive or multiplicative, this makes their application to 
collective processing much simpler, as they may be evaluated for
the individual states, and from this one obtains their values for
many copies of the same state, or for the combination of several
copies of each of several different types of state.

The monotones $E_{k_1,...,k_N}$ are supermultiplicative in the
sense given by the following proposition.
\begin{proposition}{\rm{(Supermultiplicativity)}} \label{prop: supermultiplicativity}
\beqa \label{supermultiplicativity}
E_{k_1,...,k_N}(\ket{\psi^\X}) E_{l_1,...,l_N}(\ket{\chi^\Y})
\le E_{k_1 l_1,...,k_N l_N}(\ket{\psi^\X} \odot \ket{\chi^\Y})\;.
\eeqa
\end{proposition}
The proof of this proposition is immediate from the definition of these
monotones:  one need only note that the product of a rank-$k_i$
projector $P^{\X_i}_{k_i}$ on $\X^i$ and a rank-$l_i$ projector $Q^{\Y_i}_{l_i}$
on $\Y^i$ is a rank-$k_i l_i$ projector on $\X^i \odot \Y^i$, and therefore
the value $E_{k_1,...,k_N}(\ket{\psi^\X}) E_{l_1,...,l_N}(\ket{\chi^\Y})$
is achievable in the maximization defining
$E_{k_1 l_1,...,k_N l_N}(\ket{\psi^\X} \odot \ket{\chi^\Y})$.
It is far from obvious, however, that the two quantities in the above
proposition are equal, i.e. that multiplicativity holds.
It may be that projectors which do not have a product structure with respect
to the $\X \odot \Y$ tensor factorization can achieve a higher projection.
(Werner and coworkers, however, have numerically 
investigated some cases of the maximum modulus-squared inner product
with a pure product (with respect to the party structure) state, in other
words, cases of 
$E_{1,1,...,1}$, and have not found violations of  
multiplicativity \cite{Wernercomm}.)

In some special cases, it is easy to show multiplicativity.

\begin{proposition}
For monotones $E_{k_1,...,k_N}$ and states $\ket{\psi}$ and
$\ket{\chi}$ for which 
the reduced density matrix of each state
onto the systems with less-than-full-rank $k_i$ is maximally mixed,
we {\bf do} have multiplicativity, i.e. equality in 
(\ref{supermultiplicativity}).  
\end{proposition}

\begin{proposition}
When we tensor an entangled state of all parties with
a product state $\ket{\nu}$ of all parties, the optimal projector 
may be decomposed as $Q_{k_1} \otimes \cdots \otimes Q_{k_N}
= \proj{\nu} \odot P_{k_1} \otimes \cdots \otimes P_{k_N}.$
This implies that the value of our monotones
is unchanged by tensoring
with a product pure state.
\end{proposition}
This proposition must hold for any entanglement monotone, as the 
operations of adjoining or discarding $\ket{\nu}$ are both LOCC.
In the case of our monotones, we can also use Proposition 
\ref{prop: supermultiplicativity} and the fact that the monotones are 
equal to $1$ on separable states,
plus just the fact that
adjoining $\ket{\nu}$ is LOCC.

\section{Applications of the monotones}
It seems likely that for each multipartite Hilbert space (each number
of parties $N$ and set of dimensions $d_1,...,d_N$), there exist many
fundamentally distinct types of entanglement.  This statement may
be taken in various senses.   For example, one case in which we might
wish to say that two states have fundamentally different types of 
entanglement is when
neither one can be transformed into the other with certainty via LOCC.
We say they are
incommensurable under deterministic LOCC (DLOCC, for short).
If neither one can be transformed into the other with finite probability
(i.e., the parties are not guaranteed to succeed, but there is a
finite probability of success and they know when
they have succeeded), this is a stronger sense in which the states exhibit
fundamentally different types of entanglement: we will say they 
are incommensurable under SLOCC (the S is for stochastic).  D\"ur, Vidal 
and Cirac \cite{Dur2000a}  define two states to be equivalent under SLOCC
when there is nonzero probability for a transition
in both directions (this is an equivalence relation).  (Note, however,
that the relation of incommensurability under SLOCC defined above need
not be an equivalence relation;  while symmetric, it is not obviously 
transitive.  Nor, of course, is it the complement of D\"ur et.~al's 
SLOCC-equivalence.)   Another		
sense, stronger still,  is
if there is no nonzero asymptotic rate $R$ at which $C_1$
copies of one state $\ket{\psi}$ ($\ket{\chi}$) can be LOCC-transformed 
into $C_2$ copies of
the other state $\ket{\chi} (\ket{\psi})$, 
in the limit in which $C_1,C_2 \rightarrow \infty$, 
$C_1/C_2 \rightarrow R$,
and the fidelity of 
the transformed state to the target state $\ket{\chi}^{\otimes C_2}$
($\ket{\psi}^{\otimes C_1}$)
approaches one.
The monotones $E_{\K}$ give direct information on the first two
questions, 
and if their multiplicativity properties discussed in Section \ref{collective monotones}
can be understood, or at least controlled
for some examples as $C_1,C_2 \rightarrow \infty$, then they may yield information
on the third as well.

To this end, one would like to calculate the monotones for interesting multipartite
states, hoping to find examples of pairs of states
ranked in the
reverse order by two of the $E_{\K}$.  This implies they 
are DLOCC-incommensurable.  SLOCC may be investigated using the following
observation.
\begin{proposition} 
\beqa \label{probability bound}
{\rm prob}( \ket{\psi} \rightarrow \ket{\chi}) 
\le \frac{1 - E_\K(\ket{\psi})}{1- E_{\K}(\ket{\chi})}\;.
\eeqa
\end{proposition}
\noindent
{\em Proof:}  
Suppose $\ket{\psi} \rightarrow \ket{\chi}$ via LOCC with probability
$p$.  Then $\ket{\psi}$ goes to some ensemble $\Upsilon$ containing the state 
$\sqrt{p}\ket{\chi}$. 
$E_\K(\ket{\psi}) \le E_{\K}(\Upsilon) \le (1-p) + pE_\K(\ket{\chi})$
(the first inequality is from the fact that $E$ is a monotone, and
the second is from the fact that the largest value of $E$ for an ensemble
containing $\ket{\chi}$ with probability $p$ occurs where all the other 
states are separable).  Algebra gives the proposition. \QED

This is the same argument already used by Vidal for bipartite monotones.
Note that $1-E_\K$ are {\em decreasing} entanglement monotones.
Also, the RHS  of (\ref{probability bound}) can be greater than 1 for some
monotones and states.  This
is not a problem, it merely means that the monotone in question
imposes no restriction on SLOCC transformations of the states in question.
(Other monotones may, however.)

\begin{corollary}
If $E_\K(\ket{\psi}) = 1$ while $E_\K(\ket{\chi}) \ne 1$ for some 
$E_\K$, then $\ket{\psi}$ may
not be converted into $\ket{\chi}$ via SLOCC.  
\end{corollary}

\section{Examples} \label{sec: examples}

As an example of our monotones in action, we
consider some simple tripartite states known to have irreducibly 
tripartite entanglement, the states 
$\ket{W} := 1/\sqrt{3}(\ket{001} + \ket{100} + \ket{010}$ \cite{Dur2000a}
and $\ket{\rm GHZ} := 1/\sqrt{2}(\ket{000} + \ket{111})$.  We compare
them to states with only bipartite entanglement, such as the tensor product
of a Bell state of two parties with a pure state for the third party.

\begin{itemize}
\item
GHZ state:  
$E_{2,1,1}=E_{1,2,1}=E_{1,1,2} = 1/2\; ;$
$~~~E_{1,1,1}= 1/2\;$; \\
$E_{2,1,1}~ 
(\mbox{or any permutation})~= 1/2.$
\item
Singlet for parties $1$ and $2$ tensored with a  pure state for party $3$: \\
$~~~E_{2,1,1} = E_{1,2,1} = E_{1,1,2} = 1/2\; ;$ \\
$~~~E_{1,1,1} = 1/2\;$;$~~E_{1,2,2}=E_{2,1,2}=1/2\; ;~~
E_{2,2,1}=1.$
\end{itemize}
Even taken together, the values of $E_{2,1,1}$ and
permutations, and of $E_{1,1,1}$,
do not differentiate between the GHZ and
the singlet.  Of course, 
$E_{2,2,1}$ does 
differentiate between these cases.  

Thus in the case of Bell vs. GHZ, the only one of our 
multipartite monotones which 
distinguishes the two kinds of state is essentially
bipartite in nature.  One might worry that the new multipartite
monotones never provide any interesting information about state
transformations which does not stem from bipartite considerations.
This worry would be unjustified, as a comparison of the
$\GHZ$ and $\ket{{\rm}W}$ states shows.

We have:
\begin{itemize}
\item
$\ket{W}$:  $E_{2,1,1} = 1/3\; ;$ 
$~~E_{1,1,1} = 4/9\; ;$~~
$E_{2,2,1} ({\mbox{or any permutation}}) 
= 2/3.$
\end{itemize}

$E_{2,2,1}$ tells us that
a $\ket{W}$ can't go to a $\ket{{\rm GHZ}}$ via DLOCC.  This involves only
bipartite considerations.  But $E_{1,1,1}$ tells us, in addition, that
a $\GHZ$ can't go to a $\ket{W}$.  This is not
forbidden by bipartite considerations.  So, the 
multipartite monotones we have defined
cannot all be just monotonic functions of the bipartite ones.
Rather, they give us further, irreducibly multipartite, information 
about which state transformations are possible via multipartite 
LOCC.  In fact, Proposition \ref{probability bound} 
lets us use the multipartite monotones to bound the
probabilities for conversion via SLOCC: $p \le 2/3$ (from 
$E_{2,2,1}$) for 
$\ket{W} \rightarrow \ket{{\rm GHZ}}$, $p \le 3/4$ (from 
$E_{2,1,1}$) for 
$\ket{\rm GHZ} \rightarrow \ket{W}$.

We know from 
the work of D\"ur, Vidal and Cirac \cite{Dur2000a} that
neither of these probabilities can be nonzero, so the bounds
are not tight.  So the nondecrease of
these monotones, 
even considered all together, is not a sufficient condition for
multipartite SLOCC.
This despite the fact that they
generalize, in a way that does give further irreducibly multipartite
constraints on SLOCC, a set of bipartite monotones whose nonincrease {\em is} 
known to 
be a necessary and sufficient condition even for SLOCC.

\section{Invariants}
In the rest of this paper, we introduce polynomial invariants for
simultaneous collective processing of more than one quantum state of $N$
parties.  In order to explain these invariants, it will be helpful
to first recall some information about invariants for processing
of single quantum states.

There has been much written about the situation when $N$ parties
share a single quantum state $\ket{\psi}$ and the Hilbert space of
each party is a qubit.  A useful tool for analyzing this case is
invariant theory \cite{Linden98a}.  Thus it is known that
there is an infinite set of polynomial functions of the state each
of which is invariant under local unitary transformations.  If two
$N$-party states have different values of any invariant
polynomial, then they are not transformable into each other by
local unitary transformations.  For comparison with what follows it
is useful to give an explicit example here.  So consider the case
$N=3$.  A general pure state $\ket{\psi}$ may
be written as
\begin{eqnarray}
\ket{\psi} = \sum_{i,j,k=1}^2 \psi_{ijk}\ket{i} \ket{j}\ket{k}.
\end{eqnarray}
It is useful to think of arranging the set of
polynomial invariants in order of increasing degree in the state 
$\ket{\psi}$.  Some of the invariants of low degree are
\begin{eqnarray}
I_2 &=& \sum_{i,j,k=1}^2  \psi_{ijk} \psi^*_{ijk}\label{I_2}\\
I_4^{(1)}&=& \sum_{i,j,k,m,n,p=1}^2  \psi_{ijk} \psi^*_{imn}\psi_{pmn} \psi^*_{pjk}\\
I_4^{(2)}&=& \sum_{i,j,k,m,n,p=1}^2  \psi_{jik} \psi^*_{min}\psi_{mpn}
\psi^*_{jpk}\\
I_4^{(3)}&=& \sum_{i,j,k,m,n,p=1}^2  \psi_{jki} \psi^*_{mni}\psi_{mnp}
\psi^*_{jkp}\\
I_4^{(4)}&=& \sum_{i,j,k,m,n,p=1}^2  \psi_{ijk} \psi^*_{ijk}\psi_{mnp}
\psi^*_{mnp}\\
I_6&=& \sum_{i,j,k,m,n,p=1}^2  \psi_{ijk} \psi^*_{imn}\psi_{pqn}
\psi^*_{pjs}\psi_{rms}
\psi^*_{rqk}.
\end{eqnarray}
The lower index on the invariant indicates the degree in 
$\ket{\psi}$, the upper index is used to distinguish between
invariants of the same degree. Sums and products of 
invariants are clearly also invariant (the set of
polynomial invariants forms a ring). 

The above invariants have
been constructed by contracting each local index with the
invariant tensor $\delta_{ij}$ of $U(2)$, in such a way that
a $\psi$ index is always contracted with a $\psi^*$ index; e.g. $I_2$ may be
written
\begin{eqnarray}
I_2 = \sum_{i_1,j_1,k_1, i_2,j_2,k_2=1}^2  \psi_{i_1j_1k_1}
\psi^*_{i_2j_2k_2}\delta_{i_1i_2}\delta_{j_1j_2}\delta_{k_1k_2}.
\end{eqnarray}
When an invariant is written this way, we will say it is 
{\em in simple form}.  Not all polynomial invariants can
be written in simple form;  those which can, we will call
{\em simple}.  Of course, even simple invariants can also
be written in forms which are not simple.

Another interesting invariant is the residual tangle \cite{Coffman99a}.  Its
definition is
\begin{eqnarray} \label{tangle}
T := 2 \left| \sum_{i,j,k,m,n,p,q.i',j',k',m',n',p'}
\psi_{ijk} \psi_{i'j'm} \psi_{n p k'} \psi_{n'p'm'}
\epsilon_{ii'}\epsilon_{jj'} \epsilon_{kk'} \epsilon_{mm'}
\epsilon_{nn'} \epsilon_{pp'}\right|\;.
\end{eqnarray}
where $\epsilon_{ij}$ is the antisymmetric invariant tensor of 
$SU(2)$ (the fact that we take the modulus of the
expression means that it is invariant under $U(2)$).
$T$ is not a polynomial in $\psi$ and $\psi^*$.  However its square
is a polynomial, and 
it may be written as a sum of simple terms involving
the invariant tensor $\delta$.
This may be  done by using the relation
$\epsilon_{ii'} \epsilon_{jj'} = 
\delta_{ij} \delta_{i'j'} - \delta_{ij'} \delta_{i'j}$.  If 
we pair
$\epsilon$'s from the $\psi$ terms with $\epsilon$'s from the 
$\psi^*$ terms (which are absent from within the modulus in
(\ref{tangle}),
but are 
introduced---along with twelve new indices, and the removal of 
the modulus---when (\ref{tangle}) is squared), the expression will
(when multiplied out)
be a sum of simple invariant terms, as claimed.  Explicitly:  
\begin{eqnarray}
T^2 & = & \nonumber \\ 
& & \quad 4~~\sum 
\psi_{ijk} \psi_{i'j'm} \psi_{n p k'} \psi_{n'p'm'}
\psi^*_{IJK} \psi^*_{I'J'M} \psi^*_{N P K'} \psi^*_{N'P'M'} \nonumber \\
& & \quad \quad \quad ~~~\times 
(\delta_{iI}\delta_{i'I'} - \delta_{iI'}\delta_{i'I})
(\delta_{jJ}\delta_{j'J'} - \delta_{jJ'}\delta_{j'J}) \nonumber \\
&  & \quad \quad \quad ~~~\times
(\delta_{kK}\delta_{k'K'} - \delta_{kK'}\delta_{k'K}) 
(\delta_{mM}\delta_{m'M'} - \delta_{mM'}\delta_{m'M}) \nonumber \\
&  & \quad \quad \quad ~~~\times
(\delta_{nN}\delta_{n'N'} - \delta_{nN'}\delta_{n'N}) 
(\delta_{pP}\delta_{p'P'} - \delta_{pP'}\delta_{p'P}).
\end{eqnarray}
(Possibly there are other similar expressions for 
$T^2$, which can't be matched with this one 
term-by-term by renaming dummy indices,
arising through different pairings of $\epsilon$'s when applying the 
$\epsilon-\delta$ identity.)

General theorems from invariant theory imply that
any polynomial invariant may be written, as we have done with $T^2$,
as a sum of simple
invariants.  Each simple invariant is a product of equal
numbers of $\psi$'s and $\psi^*$'s with all the local indices for
a given party contracted (pairing $\psi$-indices with 
$\psi^*$-indices) using the invariant tensor $\delta_{ij}$;
 also any such polynomial is invariant (see
\cite{Linden98a} for further details).  It is also known how
to calculate a generating function (the Molien series)
for the number of linearly
independent invariants of each degree. The simple invariants
will turn out to be particularly interesting when we come to
consider collective processing of more than one state below.  
Note however that the squared tangle is not written in simple
form above.  We believe that it cannot be written in simple form,
and thus is not simple.  

An alternative way of writing the above invariants which will be
useful below is using an \lq\lq index-free\rq\rq\ notation.  Let us write
$\ket{\psi} \bra{\psi}$ as $\rho$; we also find it helpful now to
label the parties $A,B,C$. 
Then the above invariants may be written as:
\begin{eqnarray}
I_2 &=& {\rm Tr}_{ABC} \rho \\
I_4^{(1)}&=& {\rm Tr}_{BC}
\left[ {\rm Tr}_{A} \rho\right]^2\\
I_4^{(2)}&=& {\rm Tr}_{AC}
\left[ {\rm Tr}_{B} \rho\right]^2\\
I_4^{(3)}&=&{\rm Tr}_{AB} \left[
{\rm Tr}_{C} \rho\right]^2\\
I_4^{(4)} &=&
({\rm Tr}_{ABC}\rho)^2\\
I_6&=&
{\rm Tr}_{ABC} \left[ 1_A\otimes{\rm Tr}_{A}\rho\right] \left[ 1_B\otimes{\rm Tr}_{B}\rho\right]
\left[ 1_C\otimes{\rm Tr}_{C}\rho\right].
\end{eqnarray}
In the expressions above ${\rm Tr}_{AB}$ denotes the trace over the
Hilbert space for the first and second parties and the symbol $1_A$ means the $2\times 2$
identity operator on the first Hilbert space.

Of course not all the infinitely many invariants are independent
of each other (for example $I_4^{(4)} = (I_2)^2$).  The Hilbert
basis theorem states that for any ring of invariants of interest
to us, 
there exists a finite basis of the ring (i.e. any invariant
may be written as a sum of products of elements of the basis, and
the number of elements in this basis is finite).  Thus in checking
whether two states are locally equivalent, one only needs to check
whether the elements of basis have equal values when evaluated for
the two states.  Unfortunately there is no simple
procedure for calculating the basis for any given example.  Thus
while the basis is simple to find for $N=2$ parties (in fact for
any dimension of local Hilbert space, not just qubits), and a
basis for the case $N=3$ has been reported \cite{Grasslcomm}, the
basis is not known for more parties, as far as we are aware.
Nonetheless knowing the invariants of low degree can still be very
useful in applications (see for example \cite{Kempe1999a} and below).

\sloppy
\section{Collective processing and multiplicative invariants}
\fussy
We now turn to the case that the $N$ parties share not one quantum
state but a number $\C$: 
$\{ \ket{\psi_1},\ket{\psi_2},\ldots \ket{\psi_\C}\}$, 
and we increase the types of operation the parties
are allowed to do to include unitary processing of all these
copies together.  A particularly interesting case is when all
states are copies of a single state, but most of what we will have
to say is applicable to the more general case in which the states
are different.
We will continue to use the notation $\odot,$ introduced
in Section \ref{collective monotones}, for
the tensor product of different state spaces belonging to the
same set of parties, and of states belonging to these different 
state spaces.  Recall that {\em collective} processing of states
refers to the possibility that the each party's local operations 
may be ``nonlocal'' with respect to the tensor product structure
$\odot$:  that is, the different states shared 
by the same parties may all 
be processed together, though still locally with respect to the
party structure.

In the case of bipartite systems there is a well-known function,
the local entropy $S(\rho)=-{\rm Tr}_B \left[ {\rm Tr}_A \left[\rho\right]
\log {\rm Tr}_A \left[\rho\right]\right]$ which
is invariant under collective processing; it is also additive, namely
$S(\rho_1\odot \rho_2) = S(\rho_1) + S( \rho_2)$.
Our aim here is to show that there is a
large class of polynomial functions for multi-party systems which are
invariant under collective unitary 
transformations, and which are
multiplicative (hence their logarithms, when defined, are
additive).

 The most general form of collective processing allows general
 LOCC transformations on multiple copies.  As a step towards
 understanding this general case we will discuss what can be said
 about processing with collective {\em unitary} transformations.
 To be explicit let us first consider qubit states and consider
 pure states of three parties (extensions to more parties and
 states of higher dimensional systems will then follow).  The
 space of states of single copies is $\mathbf{C}^{2}
 \otimes \mathbf{C}^{2} \otimes \mathbf{C}^{2} =  \mathbf{C}^{8}$;
 the group of local unitary transformations on this space 
is $U(2)\times U(2) \times
 U(2)$.  
  Thus the
 Hilbert space of two states is
\begin{eqnarray}
\left( \mathbf{C}^{2}\odot \mathbf{C}^{2}\right)
 \otimes \left( \mathbf{C}^{2}\odot \mathbf{C}^{2}\right)
  \otimes \left( \mathbf{C}^{2}\odot \mathbf{C}^{2}\right) =
\mathbf{C}^{4}
 \otimes \mathbf{C}^{4} \otimes \mathbf{C}^{4}.
 \end{eqnarray}
 We will be interested in properties of the combined state
 invariant under $U(4)\times U(4) \times U(4)$.

 The sort of question we would like to address is as follows.
 Kempe \cite{Kempe1999a} has produced an interesting example of two
 3-qubit states which are not locally equivalent
 although they have equal values of their
 bipartite invariants (or equivalently their local entropies).
 The non-equivalence was demonstrated by calculating the value of
 $I_6$.  The two states and the values of the invariants are
 \begin{eqnarray}
 \ket{\phi_1}{} = 1/\sqrt{37}(2 \sqrt{3} \ket{000}{} - 5 \ket{111}{} ) ; \nonumber \\
\ket{\phi_2}{} = 1/\sqrt{37}(4 \sqrt{2} \ket{000}{} - 5 \ket{+++}{} ) ; \nonumber \\
\quad I_4^{(j)}(\ket{\phi_1}{}) = I_4^{(j)}(\ket{\phi_2}{})= 769/1369~~(j=1,2,3) ;
\nonumber \\ 
I_6(\ket{\phi_1}{}) \approx 0.343 \ne I_6(\ket{\phi_2}{}) \approx 0.242\;.
 \end{eqnarray}
Here $\ket + {} := (\ket{0}{} + \ket{1}{} )/\sqrt{2}$. 
 Imagine now that one has many copies of $\ket{\phi_1}$; can they be
 transformed into many copies of $\ket{\phi_2}$ by collectively
 processing  them using unitary
 transformations?  The point is that the group we are going to
 allow to act is much larger in the collective case than in the
 case of one copy (the group of local unitary actions
 in the case of one copy is $U(2)\times U(2) \times U(2)$; in the
 case of $\C$ copies it is $U(2^\C)\times U(2^\C) \times U(2^\C)$).
 In particular, consider starting with $\C$ copies of $\ket{\phi_1}$; 
the state of these $\C$ copies lies inside $\mathbf{C}^{2^\C}
 \otimes \mathbf{C}^{2^\C} \otimes \mathbf{C}^{2^\C}$ in a
 particular way.  Using collective unitaries we can transform the
 state $(\ket{\phi_1}{})^{\otimes \C}$ into other states some of which
 may be written as the tensor power of some other state but
 the way in which the new tensor structure of each of the local Hilbert spaces
 lies inside $\mathbf{C}^{2^\C}$
  may be quite
 different from the way the Hilbert space was initially decomposed
 as a $\C$-fold tensor product.  The question is, then, does the
 extra freedom collective processing allows enable us to transform
 multiple copies of $\ket{\phi_1}{}$ into multiple copies of $\ket{\phi_2}{}$?

 We now present an infinite family of polynomials which are
 invariant under collective processing of any number of qubits.
 The easiest way to describe the family is by an example.
 Consider then a particular homogeneous polynomial of a single
 state which is invariant under $U(2)\times U(2) \times U(2)$, namely
 \begin{eqnarray}
 I_4^{(1)}= {\rm Tr}_{B_1C_1}
\left[ {\rm Tr}_{A_1}\rho\right]^2.\label{I4}
\end{eqnarray}
Here
${\rm Tr}_{A_1}$ denotes the trace of the operator over the
2-dimensional Hilbert space $A_1$ etc. (we will shortly be taking the
tensor product of more than one Hilbert space for each party); as before we use
$\rho$ to denote the density matrix associated with the pure state $\ket{\psi} \bra{\psi}$.
This invariant may be easily extended to an invariant under
collective processing of two states.  Denote by $A$ the 4-dimensional Hilbert
space of the first party etc. ; clearly
\begin{eqnarray}
 {\rm Tr}_{BC}
\left[ {\rm Tr}_{A}\rho_1 \odot \rho_2\right]^2\label{collective1}
\end{eqnarray}
is invariant under local actions of $U(4)$ by each party, since
each trace is invariant.  Of course (\ref{collective1}) is just
one of the local invariants of states of three parties each
having a four level system (see below).  Moreover for any operator $X$ on
$ \mathbf{C}^4$, we may write
\begin{eqnarray}
 {\rm Tr}_{A} (X) =  {\rm Tr}_{A_1} \left( {\rm Tr}_{A_2} \left( X
 \right) \right)
\end{eqnarray}
where ${\cal H}_A = {\cal H}_{A_1} \odot {\cal H}_{A_2}$ is the
local Hilbert space of party $A$; $A_1$ and $A_2$ label {\em
any} decomposition of the local Hilbert space ${\cal H}_A =
\mathbf{C}^4$ as a tensor product of two copies of $\mathbf{C}^2$.
If we take the decomposition corresponding to $\ket{\psi_1}$ and
$\ket{\psi_2}$ we see that
\begin{eqnarray}
 {\rm Tr}_{BC}
\left[ {\rm Tr}_{A}\rho\right]^2 &=& {\rm Tr}_{BC}
\left[ {\rm Tr}_{A_1}{\rm Tr}_{A_2}\rho_1\odot \rho_2\right]^2\nonumber\\
&=& {\rm Tr}_{BC}
\left[ {\rm Tr}_{A_1}\rho_1 \odot {\rm Tr}_{A_2} \rho_2\right]^2\nonumber\\
&=& {\rm Tr}_{BC}
\left[ {\rm Tr}_{A_1}\rho_1 \right]^2\odot \left[{\rm Tr}_{A_2} \rho_2\right]^2\nonumber\\
&=& {\rm Tr}_{{B_1}{C_1}}
\left[ {\rm Tr}_{A_1}\rho_1 \right]^2 {\rm Tr}_{{B_2}{C_2}}
\left[ {\rm Tr}_{A_2}\rho_2 \right]^2.
\end{eqnarray}
We have used the facts that $(X\odot Y)^2 = X^2 \odot Y^2$ and
${\rm Tr}_{{A_1}{A_2}}(X\odot Y) = {\rm Tr}_{{A_1}}(X ) {\rm Tr}_{{A_2}}(
Y)$.

We have thus shown how to extend the original invariant (\ref{I4}) to one
which is invariant under collective processing and also
multiplicative (so that its logarithm is additive).  We note that
it was important the invariant is simple; a sum
of invariants will not be multiplicative, in general.

An alternative way of seeing why an object like
\begin{eqnarray}
 {\rm Tr}_{BC}
\left[ {\rm Tr}_{A}\ket{\psi} 
\bra{\psi}\right]^2 \label{collective2}
\end{eqnarray}
is invariant under local unitaries and multiplicative under tensor
products is to use index notation.  We write the invariant as
\begin{eqnarray}
& & \sum_{\mu,\nu,\rho,\sigma,\tau,\eta=1}^4  \psi_{\mu \nu \rho} \psi^*_{\mu\sigma\tau}\psi_{\eta\sigma\tau}
  \psi^*_{\eta\nu\rho}\nonumber\\
  & &\quad =
 \sum_{1}^4  \psi_{\mu_1 \nu_1 \rho_1} 
     \psi^*_{\mu_2\sigma_1\tau_1}\psi_{\eta_1\sigma_2\tau_2}
  \psi^*_{\eta_2\nu_2\rho_2} \delta_{\mu_1 \mu_2}
  \delta_{\nu_1 \nu_2}\delta_{\rho_1 \rho_2}
     \delta_{\sigma_1 \sigma_2}\delta_{\tau_1 \tau_2}\delta_{\eta_1 \eta_2}
\end{eqnarray}
where $\psi_{\mu \nu \rho}$, $\mu,\nu,\rho = 1\ldots 4$ are the components of the state 
$\ket{\psi} = \ket{\psi_1} \odot \ket{\psi_2}$, and $\delta_{\mu_1 \mu_2}$ is the
invariant tensor for $U(4)$.  Under the decomposition
of each local Hilbert space as two copies of $\mathbf{C}^2$, each
index $\mu$, say, becomes a composite index $a \alpha$; $a,\alpha
= 1,2$.   The invariant tensor $\delta$ becomes
\begin{eqnarray}
  \delta_{\mu_1 \mu_2}
 = \delta_{a_1 a_2} \delta_{\alpha_1\alpha_2}.
\end{eqnarray}
The invariance of the expression (\ref{collective2}) follows from the fact
that $\delta$ is an invariant tensor for $U(4)$. A little algebra
reproduces the result that (\ref{collective2}) is multiplicative.

The latter method of proof is useful in cases where it is not easy
to see how to write a given invariant in an \lq\lq index-free\rq\rq\ way using traces.

While we have illustrated our point by examples, it should now be
clear that any simple polynomial invariant of qubits may be
extended to a multiplicative invariant for collective unitary processing
of any number of states $\ket{\psi_1}, \ket{\psi_2},\ldots$ $\ket{\psi_\C}$.
(Note, however, that if, as we believe, the squared 
tangle is not simple, this argument does not apply to it.)

The multiplicativity of these invariants, and in particular of $I_6$,
allows us to understand convertibility via collective unitaries
for the states Kempe considered.  There are no positive integers   
$C_1,C_2$ such that $C_1$ copies of $\ket{\phi_1}{}$ can be converted to
$C_2$ copies of $\ket{\phi_2}{}.$
The nontrivial but effectively 
bipartite invariants $I_4^{(j)}~(j=1,...,3),$ which 
are all equal, impose the requirement $C_1=C_2$ in order to maintain
equality of the invariants for 
${\ket{\phi_1}{}}^{\otimes C_1}$ and ${\ket{\phi_2}{}}^{\otimes C_2}$.
Thus bipartite considerations do not forbid this exact conversion, at a 
1:1 ratio.
But when $C_1=C_2$, the invariant $I_6,$  will differ for these product states
(taking values of approximately $0.343^{C_1}$ and $0.242^{C_1}$, respectively)
implying that exact unitary conversion, even collectively, is not possible
at any ratio.  Most pairs of states will be similarly constrained.

As we mentioned above, the invariants we have been thinking of as
multiplicative invariants under collective processing of qubit
states are amongst the invariants which arise when considering
local unitary actions on a single copy of an $N$-party, locally
$d$-level
system.  The Hilbert space in this latter case is
$\odot^N \mathbf{C}^d$ and the group of local unitary
transformations is $U(d)^N$.  In fact the complete set of local
unitary invariants are formed exactly as for qubits: one uses the
invariant tensor $\delta_{RS}$, $R,S = 1\ldots d$ to contract the
local indices.  General theorems from invariant theory tell us
that all polynomial invariants are sums of terms of this form 
(in fact exactly as
for qubits, only the indices run over more values).  Thus any
polynomial invariant for this case is a polynomial invariant for
collective processing of states (where the product of the
dimensions of the local Hilbert spaces is $d$).  While sums and
products of invariants are invariant, only simple polynomials
are multiplicative when thought of as invariants of collective
processing of smaller systems.

We now turn to a few remarks about local invariants for density
matrices.  \cite{Linden99a} used a particularly convenient
representation of density matrices of qubits, the Bloch
decomposition.  For example a two-qubit density matrix can be
written as
\begin{eqnarray}
\rho = {1\over 4}\left(1_2\otimes 1_2 + \alpha_i\sigma_i\otimes 1_2
+\beta_j 1_2\otimes \sigma_j + R_{ij} \sigma_i\otimes
\sigma_j\right)\label{Bloch}
\end{eqnarray}
where $i,j=1..3$, $\sigma_i$ are the Pauli matrices, and
$\alpha_i,\beta_j$ and $R_{ij}$ are real.  The action of local
unitary transformations becomes the action  of
$SO(3)\times SO(3)$ on the parameters $\alpha, \beta $ and $R$ and
this allows us to write down invariants and also find simple
generating sets for the ring of invariants (see \cite{Linden99a} for
details).  A density matrix of $d$-level systems can be
written in a similar form to (\ref{Bloch}) with the Pauli matrices
replaced by analogous $d\times d$ matrices, and the indices $i,j$
now running from $1\ldots d^2-1$.  However it is significant that
now the action of the local unitary group on the density matrix
induces the action of a subgroup of the orthogonal group on the
parameters $\alpha, \beta $ and $R$.  Thus the full set of
invariants is much more complicated when expressed as functions of
these parameters.  Therefore it is not easy to see how to
construct invariants for collective processing using this
parametrisation for density matrices.

A more fruitful approach is to write invariants in terms of local
traces, as was done for pure states above.  Indeed any of the
multiplicative invariants we described for pure states are also
invariants for density matrices (in the \lq\lq index free\rq\rq\
form, for example, it is easy to see that any invariant for pure states becomes
one for mixed states by simply replacing $\ket{\psi}\bra{\psi}
{}$ by a general density matrix $\rho$).  However there are other
invariants, of determinant type, of general density matrices.  These
arise since the local action on density matrices is essentially an
action of the special unitary group rather than full unitary
group; the special unitary group has anti-symmetric invariant
tensors.  We will return to the question of the ring of invariants
for density matrices in a future publication.

\section*{Acknowledgments}
H.~B. thanks the EU for support via the QAIP project under
IST-1999-11234, and Chris Fuchs for long ago bringing $(\ref{the goods})$ to 
his attention.  N.~L. thanks Bill Wootters for very interesting discussions.

\begin{appendix}
\section{Proof of Proposition \ref{prop: bipartite}}
Consider a pure bipartite state written in a Schmidt decomposition
with real coefficients,
\beqa
\ket{\psi^{AB}} = \sum_i \lambda_i^{1/2} \ket{i^A} \ket{i^B},
\eeqa
so that $\ket{i^A}$ and $\ket{i^B}$ are orthonormal bases.
We may write this as $(I \otimes \rho^{1/2}) \ket{\Psi},$
where $\ket{\Psi} := \sum_i \ket{i^A} \ket{i^B}$, an unnormalized
vector (note that $\ket{\Psi}$ 
depends on a choice of local bases $\ket{i^A}$ for 
$A$ and $\ket{i^B}$ for $B$).  Thus
\beqa
\bra{\psi^{AB}}  (R \otimes P)
\ket{\psi^{AB}} \nonumber \\ 
= \bra{\Psi} (I \otimes \rho^{1/2})
(I \otimes P) 
(R \otimes I)
(I \otimes \rho^{1/2})\ket{\Psi}
\nonumber \\
=
\bra{\Psi} (I \otimes \rho^{1/2})
(I \otimes P) 
(I \otimes \rho^{1/2})
(R \otimes I)
\ket{\Psi} 
\nonumber \\
= \bra{\Psi} (I \otimes \rho^{1/2})
(I \otimes P) 
(I \otimes \rho^{1/2})
( I \otimes  R^T)
\ket{\Psi} \nonumber \\
= \bra{\Psi} (I \otimes \rho^{1/2}
P \rho^{1/2} Q)
\ket{\Psi}
\eeqa

For the second equality, we just commuted operators.  The
third equality uses the identity 
(valid for any linear operator $X$)
$(X \otimes I) \ket{\Psi} = (I \otimes X^T) \ket{\Psi}$
where the transpose is done in the local basis $\ket{i^A}$
used to define $\ket{\Psi}$,
and the implicit isomorphism between $A$ and $B$ is the one that 
identifies these local bases $\ket{i^A}$ and $\ket{i^B}$ 
with each other.
(For a 
proof, cf. e.g. \cite{Jozsa94b})
The last equality 
just defines $Q := R^T$ and uses elementary tensor product manipulations.
Here, $R,P$ are projectors on spaces $A$ and $B$ respectively, with
ranks $k_A$ and $k_B$; $Q$ is a projector of rank $k_A$, by the easily checked
fact that the transpose of a projector is a projector of the same rank.
Maximizing over all projectors of these ranks is the same as holding 
$P,Q$ fixed and maximizing, 
over all unitaries $U,V$ on $A,B$ respectively, in the expression:
\beqa
\bra{\Psi} (I \otimes \rho^{1/2}
V P V^\dagger
\rho^{1/2}
U Q U^{\dagger})
\ket{\Psi}
\eeqa
Now, this is equal to 
\beqa 
\tr \rho^{1/2} V P V^\dagger \rho^{1/2} U Q U^\dagger \;.
\eeqa
And,
\beqa 
\max_{U,V}  |\tr \rho^{1/2} V P V^\dagger \rho^{1/2} U Q U^\dagger|
\le \label{what we want}
\max_{U,V,W,Y}  |\tr \rho^{1/2} V P W \rho^{1/2} U Q Y|.
\eeqa
The latter maximization is a special case of a general schema:
\beqa \label{the goods}
\max_{U,V,W,Y...Z}  |\tr A U B V C W \cdots G Z| = 
\sum_j \sigma_j(A) \sigma_j(B) \sigma_j(C) \cdots \sigma_j(G)\;,
\eeqa
which holds for any finite set of linear operators $A,B,C,...,G$ and
unitaries $U,V,W,...,Z$.  Here $\sigma_j(X)$ are the decreasingly
ordered singular values of $X$
(eigenvalues of $|X| := \sqrt{XX^\dagger}$, which are also the eigenvalues
of $\sqrt{X^\dagger X}$ although this is not generally equal to $|X|$).

For the case of two operators and two unitaries, (\ref{the goods})
was shown by von Neumann \cite{vonNeumann37a}. 
In the general case, it follows from the facts that
\beq \label{trace AU}
\max_{{\rm unitary}~U} | \tr AU| = \sum_j \sigma_j(A)\;,
\eeq
(cf. \cite{Jozsa94b}, \cite{Horn85a} Theorem 7.4.9,. p. 432),
\beqa
(\eta \preceq_w \mu) \rightarrow
\lambda \cdot \eta \preceq_w \lambda \cdot \mu\;,
\eeqa
and 
\beq \label{singular value product majorization}
\sigma(AB) \preceq_w \sigma(A) \cdot \sigma(B) \;,
\eeq
(cf. \cite{Bhatia97a}, 
Eq. IV.21, p. 94).
Here $\sigma(A)$ stands for the vector of the decreasingly 
ordered singular values of $A$, and $\preceq_w$ means ``is 
weakly majorized by,'' {\em i.e.} for all $k$ the sum of the first $k$
components of the RHS vector is greater than or equal 
to the sum of the first $k$ components of the LHS.  $\lambda, \mu$
and $\eta$ are also assumed to be vectors with decreasingly ordered
components none of which are negative, and notation such as 
$\lambda \cdot \eta$ does {\em not} stand for the usual dot 
product, but rather for the
componentwise product of two vectors, whose value is another such vector.
(Explicitly, $(\lambda \cdot \eta)_j = \lambda_j \eta_j$.)

The proof that the LHS of (\ref{the goods}) is less than or equal to
the RHS proceeds inductively as follows.
Firstly, 
\beq
\max_{U,V,W,Y...Z}  |\tr A U B V C W \cdots G Z| = 
\sum_j \sigma_j(AUBVC...G)
\eeq
by (\ref{trace AU}).  Then use of  
(\ref{singular value product majorization})
gives 
\beq
\sum_j \sigma_j(AUBVC...G)
\le \sum_j \sigma_j(AUBVC...F)\sigma_j(G)\;.
\eeq
Now $\sigma(A \ldots F) \preceq_w \sigma(AU \ldots E) \sigma(F)$ by
(\ref{singular value product majorization}), and thus
\beq
\sigma(A \ldots F) \cdot \sigma(G) \preceq_w \sigma(A \ldots F) \cdot \sigma(F) 
\cdot \sigma(G)\;.
\eeq
Repeating this step shows that
\beq
\sigma(A \ldots G) \preceq_w \sigma(A)\cdot\sigma(B) \cdots \sigma(G)\;,
\eeq
and therefore that 
\beq
\max_{U,V,W,Y...Z}  |\tr A U B V C W \cdots G Z| = 
\sum_j \sigma_j(AUBVC...G)
\le \sum_j \sigma_j(A) \cdots \sigma_j(G)\;.
\eeq

Finally we note that appropriate choice of 
the unitaries $U,V,\ldots Z$  in (\ref{the goods}) allows this 
equality to be reached.   Specifically, by 
the singular value decomposition 
they may be
chosen so as to transform, say, 
$A$ to $|A|$ and each of
$B,C,D...$ to $U|B|U^\dagger$, $V|C|V^\dagger$, 
$W|D|W^\dagger$, each of which commutes with $|A|$
and has its ordered eigenvalues associated to their common 
eigenvectors in the 
same order as $|A|$.  

Applying this result to the case at hand, we note that
the unitary freedoms on the LHS of (\ref{what we want})
are sufficient to achieve this maximum for $A = \rho^{1/2}, 
B = P, C = \rho^{1/2}, D = Q$, 
Consequently, the 
maximal value of (\ref{what we want}), which is 
$E_{k_A,k_B}(\ket{\psi^{AB}})$, 
will be the sum of the largest $\min(k_A, k_B)$
eigenvalues of $\rho.$ \QED
\end{appendix}

\bibliographystyle{IEEE}


\end{document}